\documentclass[prl,showpacs,superscriptaddress,twocolumn]{revtex4-1}

\usepackage{graphicx}% Include figure files
\usepackage{dcolumn}% Align table columns on decimal point
\usepackage{bm}% bold mat
\usepackage{amsbsy}
\usepackage{amssymb}
\usepackage{natbib}
\usepackage[thinspace,mediumqspace,textstyle,amssymb]{SIunits}

\begin{document}
\title{Room temperature $p$-induced surface ferromagnetism}
 \author{Guntram  Fischer}
 \affiliation{%
   Institut f\"ur Physik, Martin-Luther-Universit\"at
   Halle-Wittenberg, Von-Seckendorff-Platz 1, 06120 Halle, Germany
 }
 \author{Nadiezhda Sanchez}
 \affiliation{%
   Instituto de Ciencia de Materiales de Madrid,
   Consejo Superior de Investigaciones Cient{\'{\i}}ficas,
   Cantoblanco, 28049 Madrid, Spain
 }
 \author{Waheed Adeagbo}
 \affiliation{%
   Institut f\"ur Physik, Martin-Luther-Universit\"at
   Halle-Wittenberg, Von-Seckendorff-Platz 1, 06120 Halle, Germany
 }
 \author{Martin L\"uders}
 \affiliation{%
 Daresbury Laboratory, Daresbury, Warrington, WA4 4AD, UK
 }
 \author{Zdzislawa Szotek}
 \affiliation{%
 Daresbury Laboratory, Daresbury, Warrington, WA4 4AD, UK
 }
 \author{Walter M. Temmerman}
 \affiliation{%
 Daresbury Laboratory, Daresbury, Warrington, WA4 4AD, UK
 }
 \author{Arthur Ernst}
 \affiliation{%
 Max-Planck-Institut f\"ur Mikrostrukturphysik, Weinberg 2,
 06120 Halle, Germany
 }
 \author{Wolfram Hergert}
 \affiliation{%
   Institut f\"ur Physik, Martin-Luther-Universit\"at
   Halle-Wittenberg, Von-Seckendorff-Platz 1, 06120 Halle, Germany
 }
 \author{M.Carmen Mu\~noz\email{mcarmen@icmm.csic.es}}
 \affiliation{%
   Instituto de Ciencia de Materiales de Madrid,
   Consejo Superior de Investigaciones Cient{\'{\i}}ficas,
   Cantoblanco, 28049 Madrid, Spain
 }

\date{\today}
\pacs{73.20.-r, 73.22.-f, 73.23.-b}

\begin{abstract}
We prove a spontaneous magnetization of the oxygen-terminated ZnO (0001) 
surface by utilizing a multi-code, SIESTA and KKR, {\it first-principles}
approach, involving both LSDA+U and self-interaction corrections (SIC) to
treat electron  correlation effects. 
Critical temperatures are estimated from Monte Carlo simulations, showing that
at and above $\unit{300}{\kelvin}$ the surface is thermodynamically stable and
ferromagnetic. The observed half-metallicity  and long-range magnetic order
originate from the presence of {\it p}-holes in the valence 
band of the oxide. The mechanism is universal in ionic oxides and points to a 
new route for the design of ferromagnetic low dimensional systems.
\end{abstract}

\maketitle

Since the first report of magnetism in Co-doped ZnO~\cite{ueda01} spontaneous
magnetization persistent above room temperature (RT) has been found in a
number of dilute  magnetic oxides, even
for materials containing no transition-metal impurities. The presence of
ferromagnetism (FM) without magnetic elements has been revealed in thin films
and nanoparticles of undoped  HfO$_2$, ZnO, MgO, SnO$_2$, TiO$_2$ or
SrTiO$_3$~\cite{venkatesan04a,*sundaresan09,esquinazi10}, among
%SrTiO$_3$~\cite{venkatesan04a,sundaresan09,esquinazi10}, among
others. 
The measured FM in undoped, otherwise diamagnetic, bulk oxides presents several
general and distinct characteristics such as a Curie temperature ($T_{C}$) 
substantially above room temperature, small coercive fields and a similar 
dependence of the coercive field and remanent magnetization on 
temperature~\cite{esquinazi10}. This universality indicates a common origin.
However, despite the intense research the physical origin of the
magnetic ordering and, therefore, its control remains a controversial unresolved
problem~\cite{stoneham10,coey10,dietl10}. It was shown theoretically that
localized holes in bulk oxides can lead to local moments and a
half-metallic behavior~\cite{Elfimov2002}. 
However, a high $T_C$ requires defect concentrations
incompatible with the stability of the material. Furthermore, the necessary
large strength and long-range of 
the defect-defect magnetic interaction is inconsistent with the actual
localization of defect orbitals in ionic oxides~\cite{zunger10}.

In an alternative approach, which some of us
proposed a few years ago, FM is induced at the surface due to its
intrinsic characteristics: breakdown of symmetry, unsaturated bonds or
uncompensated ionic charges~\cite{gallego05,sanchez08}. The physical mechanism
underlying the spontaneous magnetization of the surface relies on the
formation of holes at the top of the oxygen derived {\it p}-like  minority valence 
band (VB) of the oxide. This mechanism is analogous to that responsible for cation
vacancies~\cite{sanvito05} or acceptor doping~\cite{pan07} induced
magnetism. Both, defects and surfaces impose boundary conditions, which give
rise to the uncompensated charges leading to magnetization. However,
surface states, although confined to the topmost layers, are usually extended
in two-dimensions~\cite{2DSTO11} and, thus, the subtle interplay between
localization and extended states, which is required for the moment formation
and strong and long-range magnetic order, can be reached more easily at
surfaces.
In addition, there is growing experimental evidence that the FM signal
is mostly concentrated in the near-surface region, grain boundaries or
nanostructure interfaces~\cite{coey08, esquinazi10,goering09}.

Here we report a multi-code {\it first-principles} study of surface induced FM 
in ZnO, considered as the prototype of a high temperature (HT) FM oxide. 
We observe a spontaneous magnetization of the O-terminated ZnO (0001) surface
when the oxygen atoms occupy the {\it fcc} three-fold coordinated sites. The 
magnetic interaction between spin moments is strong and the calculated Curie 
temperature is $\gtrsim \unit{300}{\kelvin}$.  
The surface turns out to be thermodynamically stable in the range of
oxygen pressures easily achieved experimentally. The underlying
physical mechanism is universal in ionic oxides, although the onset of the
spontaneous magnetization requires a critical number of {\it p}-holes, only
achieved in specific surfaces. 

In our calculations ZnO surfaces have been modeled by periodically repeated
slabs separated by a vacuum region. To facilitate a more adequate treatment of
electron correlations and $p$ bands containing
holes~\cite{zunger10,dps08,*0953-8984-22-43-436002}, we have employed two
%holes~\cite{zunger10,dps08,0953-8984-22-43-436002}, we have employed two
approaches. First, the LSDA+$U$ method, as implemented in the  pseudopotential
SIESTA code~\cite{siesta02}, known to well reproduce the groundstate of bulk
ZnO~\cite{supplementary}, and second, the local 
self-interaction corrections 
(LSIC), as implemented in the multiple scattering theory, the LSIC-KKR
code~\cite{LED05}. 
While the electronic structure and structural relaxations have been performed
with SIESTA, the magnetic interactions for the relaxed structures have been
investigated with the LSIC-KKR code. 
In the latter, after determining the SIC groundstate, the system is mapped 
onto a  Heisenberg Hamiltonian $H = - \sum_{ij} J_{ij} \mathbf{e}_i \cdot
\mathbf{e}_j - \sum_i \Delta_i (e_i^z)^2$ by calculating the real space
magnetic interaction parameters $J_{ij}$, where {\it i} and {\it j} stand for 
atomic sites, via the magnetic force theorem (MFT)~\cite{lka87}. $\Delta_i$ is
the magnetic anisotropy energy at site $i$, and here has been treated as a
parameter in the  performed Monte Carlo (MC) simulations to estimate Curie
temperatures~\cite{supp_large}.

\begin{figure}[htb]
  \centering
  \includegraphics[width=1.0\columnwidth]{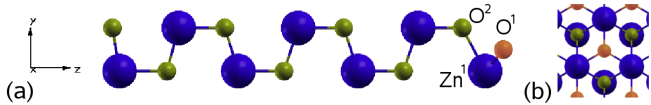}
  \caption{\label{fig:structure}
    (Color online) \textit{(a)} The investigated ZnO slab unit cell (vacuum
    region not shown) in its groundstate structure after 
    relaxation, with oxygen (small yellow and orange) and zinc (big blue)
    atoms. The ($000\bar{1}$) surface is on the left end, the ($0001$) on
    the right. The latter is characterized by  the topmost O atom in the
    three-fold coordinated hollow position, O$^1$, and the Zn and O atom
    underneath, labelled Zn$^1$ and O$^2$, respectively.    
    \textit{(b)} Topview of the ($0001$) surface.
  }
\end{figure}

Because of the symmetry of the ZnO crystal, Zn and O planes alternate along the
[0001] direction and, due to the lack of inversion symmetry 
of the wurtzite structure, the polar (0001) and ($000\overline{1}$) ZnO
surfaces are inequivalent. 
Several positions of the surface oxygen atoms have been considered, including
off-symmetric sites.
We have found that the O-ZnO ($000\overline{1}$) surface is stable and
corresponds 
simply to the cleaved ZnO bulk crystal. The VB of surface oxygen is not 
entirely filled, 
hence surface layers are metallic and exhibit partly occupied flat surface 
states at the top of the VB \cite{sanchez08,supplementary}, 
although the number of holes is small, about $0.14$ electrons. 
For the O-terminated (0001) surface, however, the most stable
configuration turns out to be the one with oxygen in the three-fold {\it fcc}
hollow site, as reported before~\cite{kress03}. This rearrangement
of the top oxygen atom decreases the surface energy dramatically by
$\approx \unit{1}{\electronvolt}$ per O atom with respect to that of the
cleaved cut-off crystal. 
The surface energy is even
about $\unit{0.65}{\electronvolt}$
smaller than the formation energy of a Zn vacancy in bulk ZnO. 
The relaxed unit cell structure is illustrated in Fig.~\ref{fig:structure}. 
The topmost oxygen atoms in both surfaces are three-fold coordinated, although
the O-Zn distances are slightly compressed and relaxed in the
($000\overline{1}$) and (0001) surfaces, respectively~\cite{supplementary}.

\begin{figure}[htb]
  \centering
  \includegraphics[width=1.0\columnwidth]{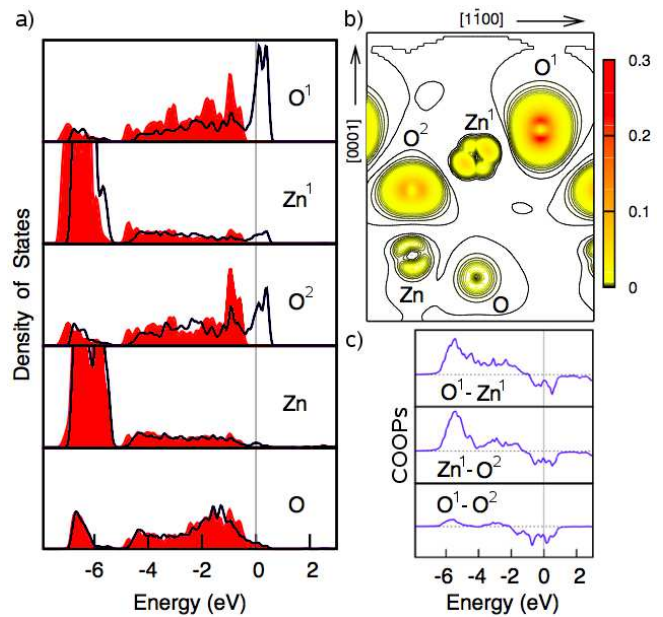}
  \caption{\label{fig:DOS_COOP}
    (Color online) \textit{a)} Spin resolved LDOS of the topmost layers of the
    O-ended (0001)-h surface. Red filled (black) values represent majority
    (minority) spin states.
    \textit{b)} Spin density distribution of the (0001)-h surface.
    \textit{c)} COOPs between the atoms in the first three layers.
    Energies are relative to $E_F$.
  }
\end{figure}

The layer resolved density of states (LDOS) corresponding to the (0001)
surface is displayed in Fig.~\ref{fig:DOS_COOP}.
The main features of the calculated LDOS reflect spin polarization and
half-metallicity of  the surface and the presence of the oxygen split-off
band, with the Fermi level crossing the minority O band. The exchange
splitting is large of 
about $\unit{1.5}{\electronvolt}$ 
and the narrowing of the minority bandwidth is particularly pronounced at the
surface layer. The total magnetic moment (MM) is 
$ \approx \unit{1.5}{\mu_B}$ and the spin densities displayed in
Fig.~\ref{fig:DOS_COOP}b confirm the surface localization of the magnetic
moments. 

The crystal orbital overlap populations (COOP) determine the wavefunction
overlap of any two atoms and, therefore, their values can be related to their
hybridization~\cite{juan03}. 
Thus, the COOP between the atoms close to the surface, shown in
Fig.~\ref{fig:DOS_COOP}c, reveal a large hybridization between adjacent layers
and evidence the extended character of states at the surface. Even the
wavefunction overlap between O$^1$ and O$^2$ is noticeable, 
which is due to their short distance, 2.30 versus % 3.25 \AA  \ 
\unit{3.25}{\angstrom} for O atoms in the same plane. 

The Mulliken charge differences and induced magnetic moments for the surface
layers are given in Table~\ref{tqmom}.
\begin{table}[htb]
  \caption{\label{tqmom}
    Layer resolved  Mulliken charge differences (in $e$) and magnetic
    moments (in $\mu_B$) for the topmost layers of the (0001) surface. The
    layers are labeled according to Fig.~\ref{fig:structure}.
  } 
  \renewcommand{\tabcolsep}{0.4pc}
  % \begin{tabular}{cp{2mm}ccp{2mm}ccc}
  \begin{ruledtabular}
  \begin{tabular}{cp{3mm}ccc}
%    \hline \hline
    & & O$^1$  & Zn$^1$ & O$^2$  \\
    \hline
    $\Delta Q$  & & -0.38  & -0.01 & -0.19  \\
    MM (SIESTA) & &  0.94  &  0.01 &  0.53  \\
    MM (LSIC-KKR)& &  1.06  & -0.06 &  0.42  
  \end{tabular}
  \end{ruledtabular}
\end{table}
It shows that there is a charge rearrangement at the topmost 
atomic layers, more pronounced in 
the surface oxygen, which sustains more than % 60\%
\unit{60}{\%}
of the induced {\it p-}holes.
The orbital charge distribution is compatible with the spatial symmetry and the
larger number of holes localized in the $p_z$ orbital~\cite{supplementary}.
The nominal Mulliken population for bulk ZnO is already reached 
three bilayers below the surface.

The surface induced {\it p-}holes give rise to an open-shell configuration 
of the oxygen which promotes the observed spontaneous spin-polarization of the
surface, analogously to the
well-known paramagnetism of the O$_2$ molecule. Therefore,
the half-metallicity of the groundstate implies that
surface magnetism is due to Hund's rule exchange, and electrons occupy
orbitals in the open shell so as to maximize their total spin.
The alignment of spins forming a symmetric spin state reduces the 
interaction energy.  Therefore,
whenever the energy gain from spin-splitting of the oxygen derived VB exceeds 
the energy loss in kinetic energy, it  
is advantageous for the system to spin-polarize. 
The COOPs displayed in Fig.~\ref{fig:DOS_COOP}c substantiate this conclusion. 
Since they represent the overlap of the 
wavefuntion of the surface atoms, they clearly show that the states have
itinerant character, and thus the delocalization
necessary for the magnetic coupling.

%%%%%%%%%%%%%%%%%%%%%%%%%%%%%%%%%%%%%%%%%%%%%%%%%%%%%%%%%%%%%%%%%%%%%%%%%%%%%
As the coordination number is reduced at the surface and the narrow oxygen
$p$ band only partially filled, it is crucial to investigate also the degree
of $p$ electron localization and its importance for an adequate description of
correlation  effects. This has been accomplished with the LSIC-KKR method, for
the relaxed surface~\cite{supplementary}. 
The global SIC energy minimum has been found for the 
scenario where, in addition to all the Zn 3$d$ electron states, all the
majority, but no  minority, {\it p} electrons of O$^1$ benefit from 
localization and hence self-interaction-correction. 
The resulting electronic structure~\cite{supplementary} confirms that the
exchange splitting is confined to the top-most oxygen sites (O$^{1}$ and
O$^{2}$) and the top of the valence band, as shown in
Fig.~\ref{fig:DOS_COOP}a. 
In Table~\ref{tqmom} we show
that the calculated MM on O$^1$ is increased due to the increased localization
of the 
majority $p$ band, causing an increased negative magnetization in the vicinity
of the surface, with the half-metallicity preserved at the surface.
The fact
that the O$^2$ majority $p$ states do not benefit from localization suggests
that its MM is induced by O$^1$ through hybridization effects.

\begin{figure}[htb]
  \centering
  \includegraphics[width=0.8\columnwidth]{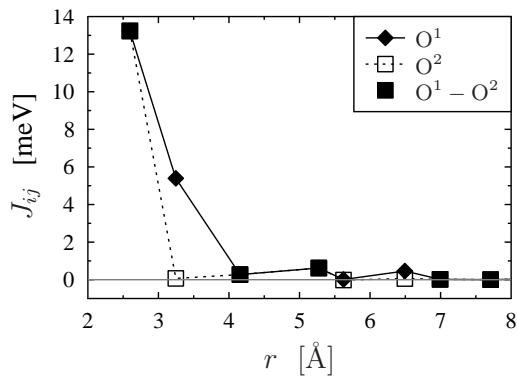}
  \caption{\label{fig:Jij}
    $J_{ij}$ of pairs containing O$^1$ ($\blacklozenge$) or O$^2$
    ($\Box$), respectively, depending on the 
    distance. The full black squares ($\blacksquare$) mark the $J_{ij}$
    between one O$^1$ and one O$^2$ atom.
  }
\end{figure}

The exchange constants $J_{ij}$ for the above SIC groundstate, calculated in the
MFT approach, are presented in % Table~\ref{tab:Jij}.
in Fig.~\ref{fig:Jij}.
First, one sees that they are strongly ferromagnetic. Furthermore,
the relevant contribution arises from interaction between
O$^1$-O$^1$ and O$^1$-O$^2$ pairs. %  up to the respective 3rd neighbor shell.
The reason for the very large nearest-neighbour (NN) coupling between O$^1$
and O$^2$ is the already mentioned close 
distance between these two atoms. The weak interaction between the O$^2$ atoms
strengthens the assumption that the MM on them is induced by O$^1$
rather than being localized there. % Table~\ref{tab:Jij}
Figure~\ref{fig:Jij} shows also that the
$J_{ij}$'s have a similar behavior to such found for other half-metallic
systems~\cite{PhysRevB.72.184415,*pssa203.2738},
%systems~\cite{PhysRevB.72.184415,pssa203.2738},
i.e.\ they oscillate and decay
relatively quickly. This means that the interaction is provided by the
delocalized electrons. 

For the MC simulations the anisotropy $\Delta_i$ has been assumed to be the
same at O$^1$ and O$^2$, and treated as a parameter. Choosing
$\Delta = 0.1$, $0.3$, $1.0$, and $\unit{3.0}{\electronvolt}$ the respective
critical temperatures have come out to be %substantially 
larger than room
temperature at $T_C= 302$, $304$, $320$, and $\unit{346}{\kelvin}$, with an 
error of $\approx \unit{6}{\kelvin}$ (\unit{15}{\kelvin} for $\Delta =
\unit{0.1}{\electronvolt}$ due to a broad transition). 

Owing to the large energy gain, the O-ZnO(0001) surface with 
O in the three-fold {\it fcc} hollow site, would be thermodynamically more
stable than the O-ZnO(0001) cleaved surface, in the 
entire range of allowed O chemical potentials. Furthermore,
the reduction of the surface energy is even sufficient to stabilize the full 
oxygen coverage of the (0001) surface for oxygen rich atmospheres  
~\cite{supplementary}.
In fact, several reports reveal the presence of oxygen atoms in ZnO(0001) 
surfaces~\cite{woll07}, either in ordered
structures~\cite{kress03,osten08,*torb09}
%structures~\cite{kress03,osten08,torb09}
or in extended O-terminated
islands~\cite{matsui07}. Therefore, the O-ZnO(0001) surface with
O in the three-fold hollow site is a good candidate to explain the
experimentally observed FM in thin layers and nanoparticles of ZnO.
Consequently, the magnetic ordering is restricted to a small 
fraction of atoms located in the vicinity of the surface, which justifies 
the low magnetization measured~\cite{coey08,esquinazi10}.
In addition, the likely presence of different polar and non-polar surfaces 
and probably adsorbates in the ZnO thin films and nanoparticles leads to a very 
complicated scenario, with magnetic and non-magnetic regions depending on the 
preparation conditions. This could explain the diversity of
magnetic behaviors found experimentally.
The possibility that boundaries are responsible for the FM of
oxides is supported experimentally, since a surface
contribution to the ferromagnetic hysteresis loops of the magnetization has 
been shown in a variety of common diamagnetic oxide
crystals~\cite{esquinazi10} and also a critical analysis of a large amount of
published data for FM in Mn doped and pure ZnO strongly suggests that grain
boundaries are the intrinsic origin for RT FM~\cite{goering09}.

Since FM is predicted in polar surfaces  
it must be related to the stabilization mechanisms of
those surfaces. Although usually their stabilization is explained in 
terms of ion removal or adsorption of charged molecules, it is still an open
question and recently there has been evidence of
electronic reconstruction and the formation of a 
2D electron gas with exotic properties at the interfaces of ionic
oxides~\cite{2DSTO11,ohtomo04,*NatMat_QHE}.
%oxides~\cite{2DSTO11,ohtomo04,NatMat_QHE}.
Therefore, spontaneous magnetization 
could be also a plausible general stabilization mechanism of polar ionic 
surfaces. The
2D electron gas induced by the uncompensated ionic charges  
at the surface can present a spin-polarized ground state.

In conclusion, by combining DFT-based and MC methods, we have predicted
ferromagnetism at the oxygen terminated polar (0001) 
surface of ZnO with a $T_C$ of $\gtrsim \unit{300}{\kelvin}$. The spontaneous
magnetization 
of the surface originates from the presence of {\it p}-holes in the minority 
valence band of the oxide, which renders a half-metallic
surface. Spin-polarization 
is enhanced through the bandwidth reduction associated to the surface, 
which drives moment formation more efficiently. In addition,
surface states, although confined to the
topmost layers, are extended in two dimensions and thus the subtle interplay
between localization and extended states required for the development 
of long-range magnetic order can be reached at the surface. The mechanism is
universal in ionic oxides, provided that the localization of the $p$
holes is strong enough for the formation of the MM.
This novel two-dimensional magnetic  
state at the surface points to a new route for the design of ferromagnetic low
dimensional systems.

\section{Acknowledgments}
This work has been partially supported by the Spanish Ministry of Science and
Technology under grant MAT2009-14578-C03-03, and by the German Research
Foundation (DFG) within the SFB~762.

\bibliography{DMO}

\end{document}